\documentstyle[prb,aps,amssymb,psfig,floats,amsmath]{revtex}

\begin{document}

\author{D. V. Dmitriev, V. Ya. Krivnov, and A. A. Ovchinnikov}
\address{Institute of Chemical Physics, Russian
Academy of Sciences, 117977~Moscow, Russia}
\title{Two-dimensional frustrated Heisenberg model: Variational study. }
\maketitle

\begin{abstract}
The stability of the ferromagnetic phase of the 2D quantum spin-$\frac 12$
model with nearest-neighbor ferro- and next-nearest neighbor
antiferromagnetic interactions is studied. It turns out that values of
exchange integrals at which the ferromagnetic state becomes unstable with
respect to a creation of one and two magnon are different. This difference
shows that the classical approximation is inapplicable to the study of the
transition from the ferromagnetic to the singlet state in contrast with 1D
case. This problem is investigated using a variational function of new type.
It is based on the boson representation of spin operators which is different
from the Holstein-Primakoff approximation. This allows us to obtain the
accurate estimate of the transition point and to study the character of the
phase transition.
\end{abstract}

\section{Introduction}

In recent years the investigation of two-dimensional frustrated Heisenberg
model is of great interest. This is mainly caused by studies of magnetic
properties of superconducting cuprates. The so called $J_1-J_2-J_3$ model
was studied by different methods in the case of completely antiferromagnetic
interactions $J_1,J_2,J_3>0$ [1-8]. An existence of disordered phases at the 
$T=0$ and nontrivial ground states in this papers is generally assumed.

The model with ferromagnetic interactions of nearest neighbors and
antiferromagnetic interactions of next nearest neighbor spins has been
investigated much less. The Hamiltonian of this model has a form:
\begin{equation}  \label{1}
H =-\sum_{{\bf n},{\bf a}}({\bf S}_{{\bf n}}\cdot {\bf S}_{{\bf n}+{\bf
a}}-\frac 14) + J\;\; \sum_{\bf n,d}
({\bf S}_{{\bf n}}\cdot {\bf S}_{{\bf n}+{\bf d}}-\frac 14) ,
\end{equation}
where vectors ${\bf a}$ and ${\bf d}$ connects nearest sites and nearest
sites along the diagonal line respectively, ${\bf n}$ - a site number and $%
J>0$. The model (1) is a special case of the $J_1-J_2-J_3$ model with $%
J_1=-1,\ J_2=J$ and $J_3=0$ ($J_3$ - is the exchange integral of the next
nearest neighbors along X and Y axes). This model is the simplest example of
the 2D frustrated system.

The ground state of the Hamiltonian (1) is ferromagnetic at small $J$ and
the energy $E=0$. But this state becomes unstable at some critical point $%
J_c $. In the classical approximation $J_c$ is equal to $1/2$ and the ground
state at $J>J_c$ corresponds to two independent sublattices with Neel order.
However, quantum fluctuations can change this situation. This question will
be discussed in this paper.

To clarify the situation let us compare the 2D model (1) with its 1D version
which has been studied in detail previously [9-11]. In the 1D model the
transition from the ferromagnetic ground state to the spiral one takes place
at $J_c=1/4$. In a recent work by two of present authors [11] the character
of this transition has been investigated by the perturbation theory in the
small parameter $(J-J_c)$ and the classical state has been used as
zeroth-order approximation. The ground state in all cases turned out to be
either with $S=0$ or with $S=S_{\max }$, and the transition from the
ferromagnetic state to the state with $S=0$ occurs by passing the states
with intermediate spins. Quantum fluctuations do not change the critical
point $J_c$, which coincides with its classical value.

\begin{figure}[t]
\unitlength1cm
\begin{picture}(11,7)
\centerline{\psfig{file=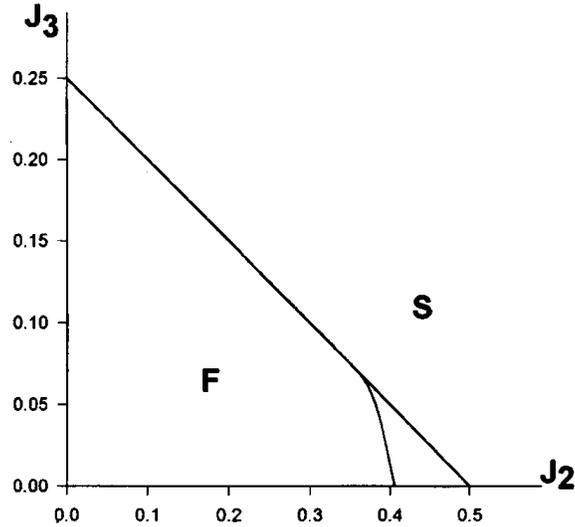,angle=0,width=9cm}}
\end{picture}
\caption[]{$T=0$ phase diagram of the  $J_1,\;J_2,\;J_3$ model with
$J_1=-1$. The bold line is a boundary between the ferromagnetic and singlet
phases in the classical approximation, the bold+thin lines -- in present
approach.}
\end{figure}

This method has been applied also to the study of the 2D model with $J_1=-1$
and $J_2,J_3>0$ in the vicinity of the phase boundary $2J_2+4J_3=1$ (Fig.1),
which determines the region of stability of the ferromagnetic state in the
classical approximation. At $J_2<0.36$ the situation is similar to the 1D
case [12]. However, at $J_2\rightarrow 0.36$ the second order of the
perturbation theory in the small parameter describing a deviation from the
phase boundary diverges. It proves that quantum fluctuations change the
classical phase diagram itself. The energies of one- and two-magnon states
are shown in Fig.2 as functions of the parameter $J$ for model (1) ($J_2=J$
and $J_3=0$ ). The critical values $J_c(S_{\max }-1)$ and $J_c(S_{\max }-2)$
of the instability of the ferromagnetic state with respect to a creation of
one and two magnons are equal to $J_c(S_{\max }-1)=1/2$ and $J_c(S_{\max
}-2)=0.4082$ respectively (two-magnon state energy was found by the
numerical solution of the corresponding Schr\"{o}dinger equation). The
difference between $J_c(S_{\max }-1)$ and $J_c(S_{\max }-2)$ shows that the
classical approach is inapplicable to the study of the stability of the
ferromagnetic state. In this respect the situation is essentially different
from those in the 1D or in the 2D cases at $J_2<0.36$, when the critical
point $J_c$ is independent on the number of magnons and coincides with its
classical value. It is natural to assume, that the critical value $%
J_c(S_{\max }-3)$ for model (1) is less than $J_c(S_{\max }-2)$, and the
critical value $J_c(0)$ corresponding to the state with $S=0$ is a true
point of the phase transition. So, the classical approximation has proved to
be unsuitable for the study of the phase transition in the model (1) and for
the determination of the critical value $J_c$ and, therefore, another
approach is needed.

\begin{figure}[t]
\unitlength1cm
\begin{picture}(11,7)
\centerline{\psfig{file=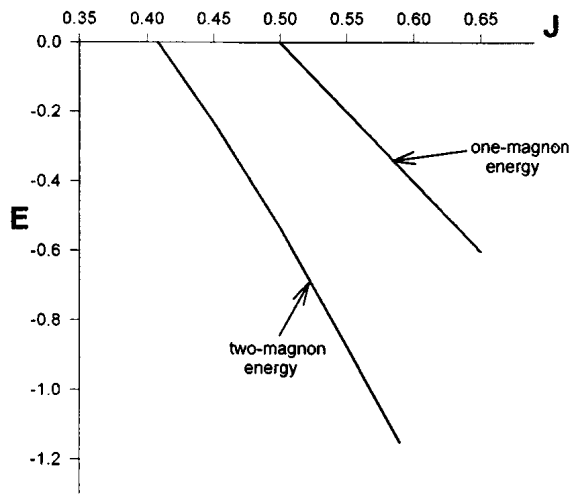,angle=0,width=9cm}}
\end{picture}
\caption[]{Energies of one- and two-magnon states.}
\end{figure}

One can obtain a crude estimate of the energy and the critical point $J_c(0)$
by using a product of the ground state wave functions of two independent
antiferromagnetic sublattices as a variational wave function (VWF). This VWF
gives for the critical point $J_c(0)$ a value:
\begin{equation}
J_c=\frac 1{1-2\varepsilon }\simeq 0.428 , \label{2}
\end{equation}
where $\varepsilon $ is the ground state energy per site of the 2D
antiferromagnetic Heisenberg model, and we use the most accurate numerical
estimations of $\varepsilon =-0.669$ [13,14]. Of course, this approach is
too poor, and the obtained value $J_c(0)$ is greater than $J_c(S_{\max }-2)$.

In the present work we propose a new type of VWF, which allows us to get
more accurate estimations of $J_c$ and to study the character of the phase
transition. This approach is based on a boson representation of the
Hamiltonian (1) which is different from the Holstein-Primakoff one. In
contrast with the spin wave theory (SWT)\ the proposed method is variational.

This article is organized as follows. In the next section we demonstrate the
method by the application it to the 2D Heisenberg antiferromagnetic model.
In section 3 the stability of the ferromagnetic phase of the frustrated
model is studied and the data are discussed.

\section{The method}

To illustrate the main features of our approach, let us consider the 2D HAF
model:
\begin{equation}  \label{3}
H = \sum_{{\bf n},{\bf a}}{\bf S}_{{\bf n}}\cdot {\bf S}_{{\bf n}+{\bf a}},
\end{equation}
where ${\bf n}$ is a site number of the 2D lattice and a vector ${\bf a}$
connects nearest sites.

It is convenient to rotate the local coordinate system of one of sublattices
by the angle $\pi $ in XZ plane:
\begin{equation}  \label{4}
H = \sum_{{\bf n},{\bf a}}(-S_{{\bf n}}^x\cdot S_{{\bf n}+{\bf a}}^x
+ S_{{\bf n}}^y\cdot S_{{\bf n}+{\bf a}}^y 
- S_{{\bf n}}^z\cdot S_{{\bf n}+{\bf a}}^z)
\end{equation}

The transformation from spin-operators to bose-ones is defined by:
\begin{equation}  \label{5}
S_i^z=-\frac{(-1)^{\stackrel{\wedge }{N}_i}}2,\qquad S_i^{+}=\frac{\theta (%
\stackrel{\wedge }{N}_i)}{\sqrt{\stackrel{\wedge }{N}_i}}b_i^{+},\qquad
S_i^{-}=b_i\frac{\theta (\stackrel{\wedge }{N}_i)}{\sqrt{\stackrel{\wedge }{N%
}_i}},
\end{equation}
where $b_i^{+}$ are bose-operators, $\stackrel{\wedge }{N}_i=b_i^{+}b_i$,
and the operator function $\theta (\stackrel{\wedge }{N})$ is:
\begin{equation}  \label{6}
\theta (\stackrel{\wedge }{N})=\frac{1-(-1)^{\stackrel{\wedge }{N}}}2
\end{equation}

It is evident that this transformation preserves all commutation relations
for the spin operators. The states with different numbers of bosons on each
site are effectively separated into equivalent unconnected pairs:
\begin{equation}
\left\{ 
\begin{array}{ccc}
S^z\left| 2m\right\rangle =-\frac 12 & ,\quad S^{+}\left| 2m\right\rangle
=\left| 2m+1\right\rangle & ,\quad S^{-}\left| 2m\right\rangle =0 \\ 
S^z\left| 2m+1\right\rangle =\frac 12 & ,\quad S^{+}\left| 2m+1\right\rangle
=0 & ,\quad S^{-}\left| 2m+1\right\rangle =\left| 2m\right\rangle
\end{array}
\right.  \label{7}
\end{equation}

As a result of the transformation (5) the Hamiltonian (4) takes the form:
\begin{equation}  \label{8}
H_b = \sum_{{\bf n},{\bf a}}\left\{ -\frac{%
(-1)^{\stackrel{\wedge }{N}_{{\bf n}}+\stackrel{\wedge }{N}_{{\bf n}+{\bf a}%
}}}4-\frac 12\left( \frac{\theta (\stackrel{\wedge }{N}_{{\bf n}})}{\sqrt{%
\stackrel{\wedge }{N}_{{\bf n}}}}b_{{\bf n}}^{+}\frac{\theta (\stackrel{%
\wedge }{N}_{{\bf n+a}})}{\sqrt{\stackrel{\wedge }{N}_{{\bf n+a}}}}b_{{\bf %
n+a}}^{+}+h.c.\right) \right\}
\end{equation}

This Hamiltonian, as well as the original one, can not be solved exactly. As
a trial wave function for Hamiltonian (8) we choose :
\begin{equation}
\left| \Psi \right\rangle =\exp \left( \frac 12\sum_{\bf i,j}%
\Lambda ({\bf i}-{\bf j)}\; b_{{\bf i}}^{+}b_{{\bf j}}^{+}\right) \left|
0_b\right\rangle , \label{9}
\end{equation}
where the function $\Lambda ({\bf i}-{\bf j)}$ will be found by the
minimization of the total energy.

The vacuum state in Eq.(9) corresponds to the state of the Hamiltonian (4)
with all spins pointing down (or corresponds to a ''chess'' arrangement of
spins for the Hamiltonian (3)).

To calculate the ground state energy we need to calculate expectation values
of all terms in the Hamiltonian (8) with respect to VWF (9). At first we
calculate the terms corresponding to the spin interactions along the
horizontal line (X axe). Owing to the translational symmetry of VWF (9), all
of these terms give equal contributions to the energy, and, therefore, it is
sufficient to calculate terms corresponding to the interaction ${\bf S}_1\cdot
{\bf S}_2$ in the original Hamiltonian (3), where 1 and 2 are nearest
neighbors along the horizontal line.

At first we represent all factors $\frac 1{\sqrt{\stackrel{\wedge }{N}_j}}$
in the Hamiltonian (8) in the form:
\begin{equation}
\frac 1{\sqrt{\stackrel{\wedge }{N}_j}}=\frac 1{\sqrt{\pi }}
\int\limits_{-\infty }^{\infty} e^{-t_j^2\stackrel{\wedge }{N}_j}dt_j
\label{10}
\end{equation}

Then the expectation value of the second term of the Hamiltonian (8),
corresponding to the term $-S_1^xS_2^x+S_1^yS_2^y$ in the Hamiltonian (4),
takes the form:
\begin{equation}
H_{12}^{xy}=-\iint\limits_{-\infty }^{\infty} \frac{dt_1dt_2%
}{8\pi }\left[ e^{-t_1^2\stackrel{\wedge }{N}_1}\left( 1-e^{i\pi \stackrel{%
\wedge }{N}_1}\right) b_1^{+}\; e^{-t_2^2\stackrel{\wedge }{N}_2}\left(
1-e^{i\pi \stackrel{\wedge }{N}_2}\right) b_2^{+}+h.c.\right]  \label{11}
\end{equation}

Thus, we need to calculate expectation values of the type:
\begin{equation}
c_{12}=\frac{\left\langle \Psi \right| \ e^{i\pi (\stackrel{\wedge }{N}_1+%
\stackrel{\wedge }{N}_2)}\ \left| \Psi \right\rangle }{\langle \Psi \mid
\Psi \rangle }\qquad  \label{12}
\end{equation}
and
\begin{equation}
\ d_{12}=\frac{\left\langle \Psi \right| \ (e^{-r_1\stackrel{\wedge }{N}%
_1}b_1^{+}e^{-r_2\stackrel{\wedge }{N}_2}b_2^{+}+h.c.)\ \left| \Psi
\right\rangle }{\langle \Psi \mid \Psi \rangle },  \label{13}
\end{equation}
where $r_1$ and $r_2$ can be:
\begin{equation}
r_j=t_j^2, \qquad \qquad r_j=t_j^2+i\pi \qquad \qquad j=1,2  \label{14}
\end{equation}

In order to decompose the boson quadratic form in Eq.(9) we use the
Hubbard-Stratanovich transformation. Now, the VWF (9) takes the form:
\begin{equation}
\left| \Psi \right\rangle =\frac{\sqrt{\det \Lambda ^{-1}}}{(2\pi )^{N_0/2}}%
\int\limits_{-\infty }^{\infty} \prod\limits_{\bf l}
d\eta_{\bf l}\; e^{-\frac 12 \sum \Lambda_{\bf ij}^{-1}\,
\eta_{\bf i}\eta_{\bf j}} \;\;
e^{\sum\eta_{\bf i}\,b_{\bf i}^{+}}\; \left| 0_b\right\rangle ,
\label{15}
\end{equation}
where $N_0$ is a number of lattice sites.

Now we can calculate the expectation values (12,13):
\begin{equation}
c_{12}=\frac{
\int\limits_{-\infty }^{\infty} \prod\limits_{\bf l}
d\xi_{\bf l}d\eta_{\bf l}\; W_{\xi \eta }\; 
e^{-2(\xi _1\eta _1+\xi _2\eta _2)}}
{\int\limits_{-\infty }^{\infty} \prod\limits_{\bf l}
d\xi_{\bf l}d\eta_{\bf l}\; W_{\xi \eta }}  , \label{16}
\end{equation}
\begin{equation}
d_{12}=\frac{
\int\limits_{-\infty }^{\infty} \prod\limits_{\bf l}
d\xi_{\bf l}d\eta_{\bf l}\; W_{\xi \eta }\; 
(\xi_1\xi _2+\eta _1\eta _2)\; e^{-r_1-r_2-p_1\xi _1\eta _1-p_2\xi _2\eta _2}}
{\int\limits_{-\infty }^{\infty} \prod\limits_{\bf l}
d\xi_{\bf l}d\eta_{\bf l}\; W_{\xi \eta }}  ,  \label{17}
\end{equation}
where we use the notations:
\begin{equation}
W_{\xi \eta }=exp \left(
-\frac 12 \sum\limits_{\bf i,j} \Lambda_{\bf ij}^{-1} \;
(\xi_{\bf i}\xi_{\bf j}+\eta_{\bf i}\eta_{\bf j})
+\sum\limits_{\bf i}\xi_{\bf i}\eta_{\bf i} \right) , \label{18}
\end{equation}
\begin{equation}
p_j=1-e^{-r_j}, \qquad \qquad j=1,2  \label{19}
\end{equation}

Using the identity:
\[
\xi _1\xi _2+\eta _1\eta _2=\frac \partial {\partial p_3}\left[ e^{p_3(\xi
_1\xi _2+\eta _1\eta _2)}\right] _{p_3=0} 
\]
Eqs.(16,17) can be written as:
\begin{equation}
c_{12}=\left\langle e^{-2(\xi _1\eta _1+\xi _2\eta _2)}\right\rangle
_{W_{\xi \eta }}  \label{20}
\end{equation}
\begin{equation}
d_{12}=e^{-r_1-r_2}\cdot \frac \partial {\partial p_3}\left[ \left\langle
T_{12}(p,\xi ,\eta )\right\rangle _{W_{\xi \eta }}\right] _{p_3=0} ,
\label{21}
\end{equation}
where
\begin{equation}
\left\langle \phi (\xi ,\eta )\right\rangle _{W_{\xi \eta }}=\frac{
\int\limits_{-\infty }^{\infty} \prod\limits_{\bf l}
d\xi_{\bf l}d\eta_{\bf l}\; W_{\xi \eta }\;  \phi (\xi ,\eta )}
{\int\limits_{-\infty }^{\infty} \prod\limits_{\bf l}
d\xi_{\bf l}d\eta_{\bf l}\; W_{\xi \eta }}  \label{22}
\end{equation}
and
\begin{equation}
T_{12}(p,\xi ,\eta )=e^{-p_1\xi _1\eta _1-p_2\xi _2\eta _2+p_3(\xi _1\xi
_2+\eta _1\eta _2)}  \label{23}
\end{equation}

It is convenient to use Hubbard-Stratanovich transformation for (23):
\begin{equation}
\left\langle T_{12}(p,\xi ,\eta )\right\rangle _{W_{\xi \eta }}=
\int\limits_{-\infty }^{\infty} \prod\limits_{j=1}^4 
dx_j \; G_{12}(p,x)\; \left\langle Q_{12}(x,\xi ,\eta )\right\rangle
_{W_{\xi \eta }} ,  \label{24}
\end{equation}
where
\begin{equation}
Q_{12}(x,\xi ,\eta )=exp\left(
\frac{ix_1+x_2}{\sqrt{2}}\xi _1+\frac{ix_1-x_2}{\sqrt{2}}\eta _1
+\frac{-ix_3+x_4}{\sqrt{2}}\xi _2+\frac{-ix_3-x_4}{\sqrt{2}}\eta _2
\right),  \label{25}
\end{equation}
\begin{equation}
G_{12}(p,x)=\frac{\exp \left( -p_2^{\prime }\frac{x_1^2+x_2^2}2-p_1^{\prime
}\frac{x_3^2+x_4^2}2+p_3^{\prime }(x_1x_3+x_2x_4)\right) }{(2\pi )^2\,
(p_1p_2-p_3^2)}  \label{26}
\end{equation}
and
\[
p_j^{\prime }=\frac{p_j}{p_1\,p_2-p_3^2} 
\]

So, there are only linear terms on $\xi $ and $\eta $ in exponent of
Eq.(25), and, therefore, we can compute the expectation value $\left\langle
Q_{12}(x,\xi ,\eta )\right\rangle_{W_{\xi\eta}}$ by diagonalizing $W_{\xi\eta}$.

Using Fourier transformation for $W_{\xi \eta }$ , we obtain:
\begin{equation}
W_{\xi \eta }=\exp \left( -\frac 12 \sum\limits_{\bf k}\left[ \omega _{%
{\bf k}}(\xi _{{\bf k}}\xi _{-{\bf k}}+\eta _{{\bf k}}\eta _{-{\bf k}})-(\xi
_{{\bf k}}\eta _{-{\bf k}}+\xi _{-{\bf k}}\eta _{{\bf k}})\right] \right) ,
\label{27}
\end{equation}
where $\omega _{{\bf k}}$ are eigenvalues of the matrix $\Lambda _{{\bf ij}%
}^{-1}$, and ${\bf k=(}k_x,k_y)$.

Diagonalizing the quadratic form in Eq.(27), we find:
\begin{equation}
\left\langle Q_{12}(x,\xi ,\eta )\right\rangle _{W_{\xi \eta }}
= exp \left( -\frac{%
x_1^2+x_3^2}2\sigma ^{+}-\frac{x_2^2+x_4^2}2\sigma ^{-}+x_1x_3\mu
_{1-2}^{+}-x_2x_4\mu _{1-2}^{-}\right) ,  \label{28}
\end{equation}
where we use the notation:
\begin{eqnarray}
\sigma ^{\pm } & = & \pm \frac 1{N_0}\sum\limits_{\bf k}\frac 1{\omega _{\bf k}\mp
1}, \nonumber  \\ 
\mu _{{\bf n}-{\bf m}}^{\pm } & = & \pm \frac 1{N_0}\sum\limits_{\bf k}\frac{%
\cos({\bf k}\cdot{\bf n}-{\bf k}\cdot{\bf m})}{\omega_{\bf k}\mp 1}
\label{29}
\end{eqnarray}

Substituting (28)\ into Eq.(24), we arrive to:
\begin{equation}
\left\langle T_{12}(p,\xi ,\eta )\right\rangle _{W_{\xi \eta }}
=\iint\limits_{-\infty }^{\infty } \prod\limits_{j=1}^4 dx_j
\frac{ exp\left(-\frac 12 \sum B_{ij}\,x_i\,x_j \right)}
{(2\pi )^2\, (p_1p_2-p_3^2)}
=\frac{\sqrt{\det B^{-1}}}{p_1p_2-p_3^2} ,  \label{30}
\end{equation}
where matrix $B_{ij}$ has the form:
\begin{equation}
B_{ij}=\left( 
\begin{array}{cccc}
p_2^{\prime }+\sigma ^{+} & 0 & -p_3^{\prime }-\mu _{1-2}^{+} & 0 \\ 
0 & p_2^{\prime }+\sigma ^{-} & 0 & -p_3^{\prime }+\mu _{1-2}^{-} \\ 
-p_3^{\prime }-\mu _{1-2}^{+} & 0 & p_1^{\prime }+\sigma ^{+} & 0 \\ 
0 & -p_3^{\prime }+\mu _{1-2}^{-} & 0 & p_1^{\prime }+\sigma ^{-}
\end{array}
\right)  \label{31}
\end{equation}

Substituting Eqs.(30,31) into (21),\ we find:
\begin{equation}
d_{12}=e^{-r_1-r_2}\left( \mu _{1-2}^{+}\; f_{+}^3(p_1,p_2)\;
f_{-}(p_1,p_2)-\mu _{1-2}^{-}\; f_{+}(p_1,p_2)\;
f_{-}^3(p_1,p_2)\right) , \label{32}
\end{equation}
where
\begin{equation}
f_{\pm }(p_1,p_2)=\left[ (1+p_1\sigma ^{\pm })(1+p_2\sigma ^{\pm
})-p_1p_2(\mu _{1-2}^{\pm })^2\right] ^{-\frac 12}  \label{33}
\end{equation}

There are four terms in (11), corresponding to the different values $r_1$, $%
r_2$ and $p_1$, $p_2$ in (14) and (19).
\begin{equation}
\left\{ 
\begin{array}{c}
p_\alpha =1+\alpha e^{-t_1^2} \\ 
q_\beta =1+\beta e^{-t_2^2}
\end{array}
\right. \qquad \qquad \alpha ,\beta =\pm 1  \label{34}
\end{equation}

Using the notation (34), Eq.(32) and the definition of functions $f_{+}$ and 
$f_{-}$ in (33), Eq.(11) can be written as:
\begin{equation}
E_{12}^{xy}=-E_{12}^x+E_{12}^y ,  \label{35}
\end{equation}
\begin{equation}
E_{12}^x=\mu _{1-2}^{+} \iint\limits_{-\infty }^{\infty }
\frac{dt_1dt_2}{8\pi }e^{-t_1^2-t_2^2} \sum\limits_{\alpha ,\beta =-1}^1
f_{+}^3(p_\alpha ,q_\beta )\; f_{-}(p_\alpha ,q_\beta) ,  \label{36}
\end{equation}
\begin{equation}
E_{12}^y=\mu_{1-2}^{-} \iint\limits_{-\infty }^{\infty }
\frac{dt_1dt_2}{8\pi }e^{-t_1^2-t_2^2} \sum\limits_{\alpha ,\beta =-1}^1
f_{+}(p_\alpha ,q_\beta )\; f_{-}^3(p_\alpha ,q_\beta)  \label{37}
\end{equation}

We note, that the expectation value $c_{12}$ can be obtained from Eq.(30) by
setting $p_1=p_2=2$ and $p_3=0$. Hence, we obtain:
\begin{equation}
E_{12}^z=\frac 14 c_{12}=\frac 14 f_{+}(2,2)\, f_{-}(2,2)
\label{38}
\end{equation}

The contribution to the energy of the terms corresponding to nearest
neighbors along the horizontal line in the Hamiltonian (8) is:
\begin{equation}
E_{12}=-E_{12}^x+E_{12}^y-E_{12}^z  \label{39}
\end{equation}

One can verify by rotating the local coordinate systems and repeating the
calculations (5-30) for the transformed Hamiltonian, that the values $%
E_{12}^x$ and $E_{12}^y$ correspond to expectation values :
\begin{equation}
E_{12}^x=\left\langle S_1^xS_2^x\right\rangle , \qquad \qquad
E_{12}^y=\left\langle S_1^yS_2^y\right\rangle  \label{40}
\end{equation}

The calculation of the expectation values of the terms of the Hamiltonian
(8) corresponding to the nearest neighbor interactions along the vertical
line (Y axe) ${\bf S}_1\cdot{\bf S}_3$ can be carried out in the analogy with
(9-30). In this case, we must only substitute the functions $\mu _{1-3}^{\pm
}$ for $\mu _{1-2}^{\pm }$ into Eqs.(36-38). Then we obtain the contribution
to energy of the vertical neighbor terms:
\begin{equation}
E_{13}=-E_{13}^x+E_{13}^y-E_{13}^z  \label{41}
\end{equation}

The total energy of Hamiltonian (8) can be written as:
\begin{equation}
E=E_{12}+E_{13}  \label{42}
\end{equation}

Thus, we find energy as a function of parameters $\sigma ^{\pm },\mu
_{1-2}^{\pm },\mu _{1-3}^{\pm }$. These parameters are functions of $\omega (%
{\bf k)}$ [or of $\Lambda ({\bf i}-{\bf j})$ in VWF (9)]. Now we need to
minimize the energy with respect to $\omega ({\bf k)}$:
\begin{equation}
\frac{\partial E}{\partial \omega _{{\bf k}}}=-\frac{a_1+b_1\cos k_x+c_1\cos
k_y}{(\omega _{{\bf k}}-1)^2}+\frac{a_2+b_2\cos k_x+c_2\cos k_y}{(\omega _{%
{\bf k}}+1)^2}=0 , \label{43}
\end{equation}
where
\[
\begin{array}{ccc}
a_1=\frac{\partial E}{\partial \sigma ^{+}}, & \qquad b_1=\frac{\partial E}{%
\partial \mu _{1-2}^{+}}, & \qquad c_1=\frac{\partial E}{\partial \mu
_{1-3}^{+}}, \\ 
a_2=\frac{\partial E}{\partial \sigma ^{-}}, & \qquad b_2=\frac{\partial E}{%
\partial \mu _{1-2}^{-}}, & \qquad c_2=\frac{\partial E}{\partial \mu
_{1-3}^{-}}
\end{array}
\]

We can define the functional form of $\omega ({\bf k)}$ from Eq.(43):
\begin{equation}
\omega ({\bf k)=}\frac{1+g({\bf k})}{1-g({\bf k})} ,  \label{44}
\end{equation}
where
\begin{equation}
g({\bf k})=\alpha _1\sqrt{\frac{1+\alpha _2\cos k_x+\alpha _3\cos k_y}{%
1+\alpha _4\cos k_x+\alpha _5\cos k_y}} , \label{45}
\end{equation}
and $\alpha _i$ $(i=1,5)$ are variational parameters.

Thus, we reduce the problem of the energy minimization over the variational
function $\omega ({\bf k)}$ to the energy minimization with respect to five
variational parameters $\alpha _i$. This procedure has been performed
numerically. It gives the final result for 2D HAF:
\begin{equation}
\alpha _1=1 ,\qquad \alpha _2=\alpha _3=-\alpha _4=-\alpha _5=-0.445 ,\qquad
E=-0.640  \label{46}
\end{equation}

Besides, both terms in Eq.(42) give equal contributions to the energy:
\[
E_{12}=E_{13} 
\]

Obtained ground state energy for the 2D HAF (3) is 4-4,5\% higher than the
most accurate results obtained by various numerical techniques [13,14] and
is in a good agreement with energies obtained by different VWF methods
[15-17].

\section{Results and discussion}

Now we apply the proposed approach to the frustrated model (1). The ground
state of the Hamiltonian (1) is ferromagnetic at small $J$. In the classical
approximation two-sublattice Neel state is realized at $J>\frac 12$. The
energy of this state is:
\begin{equation}
\frac E{N_0}=-(J-\frac 12)  \label{47}
\end{equation}

A spiral state (the incommensurate phase with the momentum $(Q,Q)$, $\cos
Q=\frac 1{2\cdot J}$) has higher energy at $J>\frac 12$:
\begin{equation}
\frac E{N_0}\simeq -(J-\frac 12)^2  \label{48}
\end{equation}
although it tends to zero at $J\rightarrow \frac 12$ as well.

Energies of other phases are considerably higher. The energy of the dimer
phase, for example, equals to $\frac E{N_0}=\frac 12$ at $J=\frac 12$.
Hence, it is natural to consider Neel two-sublattice and the spiral states
as main candidates for the ground state at $J>J_c$. We will calculate
quantum corrections to the classical energy for these states using the VWF
(9). Before that, we make the following remark.

The VWF (9) has no rotational symmetry ( it is not the eigenfunction of $S^2$%
), and, therefore, the ground state energy calculated in this approximation
depends on the choice of the local coordinate system for the Hamiltonian
(1). In other words, if we rotate spin operators ${\bf S}\rightarrow {\bf %
\varsigma }$ and perform a transformation to bose-operators according to
Eq.(5), then, generally speaking, obtained energies will be different in
spite of the equivalence of the Hamiltonians $H_{{\bf s}}$ and $H_{{\bf %
\varsigma }}$. Hence, parameters of the rotation can be considered as
variational ones.

For the consideration of the two-sublattice Neel phase let us rotate the
coordinate system so that there are the angle $\pi $ in $XZ$ plane between
two nearest neighbors on each sublattice (as in the 2D\ HAF case) and the
angle $\varphi $ between sublattices in $XZ$ plane. We will keep the angle $%
\varphi $ as a variational parameter (we note, that the energy is infinitely
degenerated with respect to $\varphi \,$in the classical approximation). In
this case, the Hamiltonian (1) takes the form:
\begin{eqnarray}
H = &-&\sum_{\bf n} [-\frac 12+\cos \varphi
\left( \varsigma _{{\bf n}}^x\varsigma _{{\bf n}+{\bf a}_x}^x+\varsigma _{
{\bf n}}^z\varsigma _{{\bf n}+{\bf a}_x}^z-\varsigma _{{\bf n}}^x\varsigma _{
{\bf n}+{\bf a}_z}^x-\varsigma _{{\bf n}}^z\varsigma _{{\bf n+a}_z}^z\right)
\nonumber \\ 
&+&\sin \varphi \left( \varsigma _{{\bf n}}^z\varsigma _{{\bf n}+{\bf a}
_x}^x-\varsigma _{{\bf n}}^x\varsigma _{{\bf n}+{\bf a}_x}^z-\varsigma _{
{\bf n}}^z\varsigma _{{\bf n+a}_z}^x+\varsigma _{{\bf n}}^x\varsigma _{{\bf
n+a}_z}^z\right) +\varsigma _{{\bf n}}^y\varsigma _{{\bf n}+{\bf a}
_x}^y+\varsigma _{{\bf n}}^y\varsigma _{{\bf n+a}_z}^y]
\nonumber \\ 
&+&J \sum_{\bf n,d}\left( -\varsigma _{{\bf n}}^x\varsigma _{{\bf %
n}+{\bf d}}^x+\varsigma _{{\bf n}}^y\varsigma _{{\bf n}+{\bf d}}^y-\varsigma
_{{\bf n}}^z\varsigma _{{\bf n}+{\bf d}}^z-\frac 14\right) ,  \label{49}
\end{eqnarray}
where vector ${\bf d}$ connects nearest sites on diagonal line.

Expectation values of the terms in the Hamiltonian (49), such as $%
\left\langle {\bf \varsigma }_{{\bf n}}^x\cdot {\bf \varsigma }_{{\bf m}%
}^x\right\rangle $, $\left\langle {\bf \varsigma }_{{\bf n}}^y\cdot {\bf %
\varsigma }_{{\bf m}}^y\right\rangle $ and $\left\langle {\bf \varsigma }_{%
{\bf n}}^z\cdot {\bf \varsigma }_{{\bf m}}^z\right\rangle $ are determined
by Eqs.(36,37,38) respectively, where $\mu _{{\bf n}-{\bf m}}^{+}$ and $\mu
_{{\bf n}-{\bf m}}^{-}$ are defined by Eq.(29).

The VWF (9) contains only terms that involve even numbers of boson operators
and, therefore, all expectation values such as $\left\langle {\bf \varsigma }%
_{{\bf n}}^x\cdot {\bf \varsigma }_{{\bf m}}^z\right\rangle $ in (49) are
equal to zero:
\begin{equation}
\left\langle {\bf \varsigma }_{{\bf n}}^x\cdot {\bf \varsigma }_{{\bf m}%
}^z\right\rangle =0  \label{50}
\end{equation}

In this case there are nine variational parameters in Eq.(45):
\begin{equation}
g({\bf k})=\alpha _1\sqrt{\frac{1+\alpha _2\cos k_x+\alpha _3\cos k_z+\alpha
_4\cos (k_x+k_z)+\alpha _5\cos (k_x-k_z)}{1+\alpha _6\cos k_x+\alpha _7\cos
k_z+\alpha _8\cos (k_x+k_z)+\alpha _9\cos (k_x-k_z)}}  \label{51}
\end{equation}

\begin{figure}[t]
\unitlength1cm
\begin{picture}(11,8)
\centerline{\psfig{file=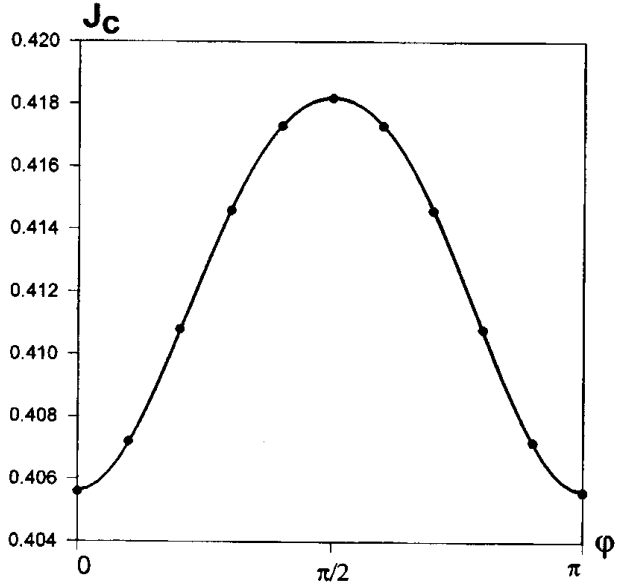,angle=0,width=10cm}}
\end{picture}
\caption[]{The dependence $J_c$ on the angle $\varphi$.}
\end{figure}

The critical value $J_c$ is defined by the condition that the ground state
energy is negative for $J>J_c$. To find $J_c$ we minimize the following
ratio:
\begin{equation}
J_c=\frac{E_1}{E_2} , \label{52}
\end{equation}
where $E_1$ and $E_2$ are the contributions to the energy from the first and
the second terms in the Hamiltonian (49):
\begin{equation}
E=-E_1+J\, E_2  \label{53}
\end{equation}

The dependence of $J_c$ on $\varphi $ obtained by a minimization with
respect to parameters $\alpha _i$ is shown on Fig.3. As it can be seen from
Fig.3, the minima of $J_c$ are reached at $\varphi =0$ or $\varphi =\pi $.
The corresponding states belong to the so called collinear phase. At $%
\varphi =0$ we have:
\begin{eqnarray}
&&\alpha _1=1 ,\qquad \alpha _2=\alpha _6=-0.7 , \qquad \alpha _3=-\alpha _7=0.64,
\nonumber \\ 
&&\alpha _4=\alpha _5=-\alpha _8=-\alpha _9=-0.36 , \qquad J_c=0.4056
\label{54}
\end{eqnarray}

The calculation of the energy with use of the VWF (9)\ for the spiral phase
can be carried out in a similar way. It turns out that quantum fluctuations
do not shift the transition point and change the coefficient in the
quadratic dependence in (48) only. The dependences of the energies of the
collinear and spiral states are shown on Fig.4. For comparison, we show the
spin wave theory results as well. The SWT energy is not a variational one
and is defined up to $J=0.52$, where the sublattice magnetization vanishes
in the SWT approximation.

\begin{figure}[t]
\unitlength1cm
\begin{picture}(11,8)
\centerline{\psfig{file=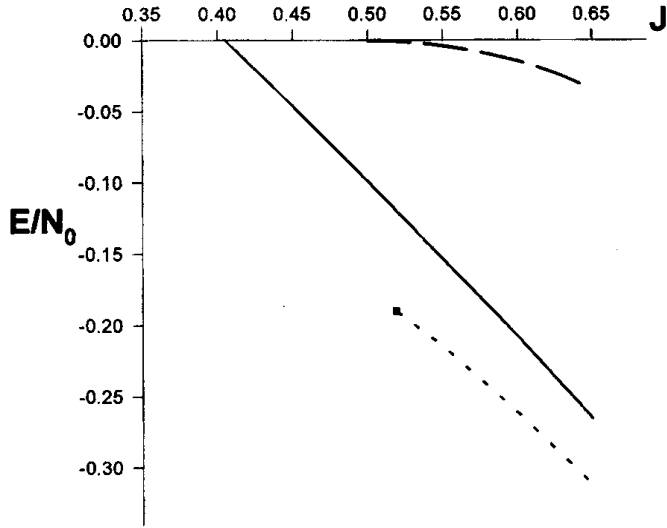,angle=0,width=10cm}}
\end{picture}
\caption[]{Energies of the collinear (solid line) and spiral (dashed)
states. Dotted line is the SWT energy of the collinear phase.}
\end{figure}

Thus, the singlet collinear phase is the ground state of the model (1) for $%
J>J_c$.

A characteristic feature of the considered model (1) is the fact, that the
critical value $J_c$, (the point of the instability of the ferromagnetic
state), depends on $S$ and $J_c(S)<J_c(S+1)$. In this respect this model
differs from the 1D case where $J_c$ does not depend on $S$. Such 1D
behaviour takes place also in a more general 2D $J_2-J_3$ model in the
vicinity of the classical phase boundary and at $J_2<0.36$. In this case,
quantum effects do not change the classical boundary of the instability of
the ferromagnetic state. The region near the boundary can be considered in
the framework of the perturbation theory, and the transition from the
ferromagnetic state to the spiral one occurs in this case. The character of
the transition changes at $J_2>0.36$, and the critical parameter $J_c$,
corresponding to $E(S)=0$, depends on $S$. In this case the transition from
the ferromagnetic to the collinear state is realized. The boundary of the
stability of the ferromagnetic phase has a form shown on Fig.1.

In conclusion, we have studied the transition from the ferromagnetic to the
singlet state in the 2D frustrated spin model. The transition region is
characterized by strong quantum fluctuations and can not be described by the
classical approximation. To study the behaviour of the system close to the
ferromagnetic boundary we have proposed new approach based on the
bozonization of the spin operators. This approach is different from the
Holstein-Primakoff method and is variational. We believe that the proposed
method can be used to the study of other Heisenberg models with the
frustration.

\section{Acknowledgments}

The authors are grateful to Prof. M.Ya.Ovchinnikova for stimulating
discussions. This work is supported by the ISTC under Grant No.015 and in
part by RFFR under Grant No.96-03-32186.


\begin{thebibliography}{99}
\bibitem{1}  A.Moreo, E.Dagotto, T.Jolicoeur and J.Riera, Phys.Rev.B 42
(1990) 6238

\bibitem{2}  P.Locher, Phys.Rev.B 41 (1990) 2537

\bibitem{3}  E.Dagotto and A.Moreo, Phys.Rev.B 39 (1989) 4744

\bibitem{4}  P.Chandra and B.Doucot, Phys.Rev.B 38 (1988) 9335

\bibitem{5}  A.V.Chubukov and T.Jolicoeur, Phys.Rev.B 44 (1991) 12050

\bibitem{6}  P.Chandra, P.Coleman and A.I.Larkin, J.Phys.Condens.Matter 2
(1990) 7933

\bibitem{7}  D.J.J.Farnell and J.B.Parkinson, J.Phys.Condens.Matter 6 (1994)
5521

\bibitem{8}  E.Dagotto, Int.J.Mod.Phys.B5 (1991) 907

\bibitem{9}  A.V.Chubukov, Phys.Rev.B 44 (1991) 12050

\bibitem{10}  R.Bursill, G.A.Gering, D.J.J.Farnell, J.B.Parkinson, Tao
Xiang, Chen Zeng, J.Phys.Condens.Matter 7 (1995) 8605

\bibitem{11}  V.Ya.Krivnov, A.A.Ovchinnikov, Phys. Rev.B 53 (1996) 6435

\bibitem{12}  D.V.Dmitriev, V.Ya.Krivnov, A.A.Ovchinnikov, Physics Letters A
207 (1995) 385

\bibitem{13}  R.A.Sauerwein, M.J.de Oliveira, Phys. Rev. B49, (1994) 5983

\bibitem{14}  J.Carlson, Phys. Rev. B40, (1989) 846

\bibitem{15}  D.A.Huse, V.Elser, Phys. Rev. Lett. 60, (1988) 2531

\bibitem{16}  W.von der Linden, M.Ziegler, P.Horsch, Phys. Rev. B40, (1989)
7435

\bibitem{17}  V.Ya.Krivnov, V.N.Likhachev, A.A.Ovchinnikov, Phys. Lett.
A192, (1994) 425
\end{thebibliography}
\end{document}